\documentclass{article}

\usepackage[preprint]{neurips_2024}
\usepackage[table]{xcolor}
\usepackage[breakable]{tcolorbox}
\usepackage{amsmath}
\usepackage{colortbl}
\usepackage{times}

\usepackage[utf8]{inputenc} 
\usepackage[T1]{fontenc}    
\usepackage{hyperref}       
\usepackage{url}            
\usepackage{booktabs}       
\usepackage{amsfonts}       
\usepackage{nicefrac}       
\usepackage{microtype}      
\usepackage{xcolor}         
\usepackage{pifont} 
\usepackage{makecell}
\usepackage{fontawesome}

\usepackage{graphicx} 
\usepackage{multirow}
\usepackage{graphicx}
\usepackage{amsmath}
\usepackage{wrapfig}
\usepackage{bm}
\definecolor{lightRed}{RGB}{255,128,128}
\definecolor{lightYellow}{RGB}{255,255,128}
\definecolor{lightGreen}{RGB}{128,255,128}
\newcommand{\dataset}[1]{\text{AJailBench}}
\newcommand{\method}[1]{\text{BoN}}

\title{Audio Jailbreak: An Open Comprehensive Benchmark for Jailbreaking  Large Audio-Language Models}

\author{
Zirui Song$^{1}$\thanks{Equal Contribution} \ \ \  
Qian Jiang $^{1*}$  \ \ \  
Mingxuan Cui$^{1*}$ \ \ \ 
Mingzhe Li$^{2}$ \ \ \ 
Lang Gao $^{1}$ \ \ \
\textbf{Zeyu Zhang} $^{3}$ \\     
\textbf{Zixiang Xu} $^{1}$ \ \ \
\textbf{Yanbo Wang} $^{1}$ \ \ \ 
\textbf{Chenxi Wang} $^{1}$ \ \ \ 
\textbf{Guangxian Ouyang} $^{1}$ \ \ \
\textbf{Zhenhao Chen} $^{1}$ \\
\textbf{Xiuying Chen} $^{1}$\thanks{Corresponding Author.} 
\\
\textsuperscript{1} Mohamed bin Zayed University of Artificial Intelligence \\
\textsuperscript{2} ByteDance
\textsuperscript{3} Australia National University
}

\begin{document}

\maketitle

\begin{abstract}

The rise of Large Audio-Language Models (LAMs) brings both potential and risks, as their audio outputs may contain harmful or unethical content. 
However, current research lacks a systematic, quantitative evaluation of LAM safety—especially against jailbreak attacks, which are challenging due to the temporal and semantic nature of speech.
To bridge this gap, we introduce \textbf{AJailBench}, the first benchmark specifically designed to evaluate jailbreak vulnerabilities in LAMs. 
We begin by constructing \textit{AJailBench-Base}, a dataset of 1,495 adversarial audio prompts spanning 10 policy-violating categories, converted from textual jailbreak attacks using realistic text-to-speech synthesis. 
Using this dataset, we evaluate several state-of-the-art LAMs and reveal that none exhibit consistent robustness across attacks.
To further strengthen jailbreak testing and simulate more realistic attack conditions, we propose a method to generate dynamic adversarial variants. Our Audio Perturbation Toolkit (APT) applies targeted distortions across time, frequency, and amplitude domains. To preserve the original jailbreak intent, we enforce a semantic consistency constraint, and employ Bayesian optimization to efficiently search for perturbations that are both subtle and highly effective.
This results in \textit{AJailBench-APT+}, an extended dataset of optimized adversarial audio samples.
Our findings demonstrate that even small, semantically preserved perturbations can significantly reduce the safety performance of leading LAMs, underscoring the need for more robust and semantically aware defense mechanisms.
We release AJailBench, including both static and optimized adversarial data, to facilitate future research: \url{https://github.com/mbzuai-nlp/AudioJailbreak}
 
\textcolor{red}{\textbf{Warning: This paper contains examples of harmful language. Reader discretion is recommended.}}
\label{sec:abstract}

%
 
\end{abstract}

\section{Introduction}
The concept of artificial assistants, a long-standing staple of science fiction, is increasingly becoming a reality in the field of artificial intelligence. Recently, the development of Large Language Models (LLMs) has seen their deployment across various domains, including virtual agents~\citep{li2024appagent,song2024mmac,liu2024tiny}, embodied robots~\citep{song2024hazards}, and medical diagnosis~\citep{han2024medinst,xie2025medtrinity25mlargescalemultimodaldataset}. 
Extending this progress, LAMs are further narrowing the gap between fiction and reality~\citep{deshmukh2023pengi,nachmani2023spoken,wang2023viola,ghosh2024gama,speechteam2024funaudiollm,gong2023listen,tang2023salmonn,wu2023next,zhang2023speechgpt,chu2023qwen,fang2024llama,xie2024mini}.
With OpenAI’s GPT-4 enabling scheduled tasks, voice-interactive AI assistants such as those imagined in science fiction are becoming possible, allowing users to perform actions like making phone calls, sending emails, and setting reminders through voice.
It is therefore critical to ensure that LAMs are aligned with safety standards to prevent the generation of harmful or unethical responses.

However, most existing research focuses on the vulnerabilities of LLMs and Large Vision Models (LVMs) under jailbreak attacks, while studies targeting LAMs remain significantly limited.
Some prior works~\citep{ying2024unveiling, shen2024voice} merely convert textual jailbreak benchmarks like AdvBench into speech form and manually test the jailbreak capabilities of GPT-4o’s audio modality. 
\textcolor{black}{These approaches are relatively naive, primarily focusing on semantic-level attacks while overlooking the unique acoustic characteristics and perturbation space of the audio modality. As a result, they fall short in comprehensively evaluating the safety robustness of LAMs.}

\begin{figure*}[t]
\centering
\includegraphics[scale=0.43]{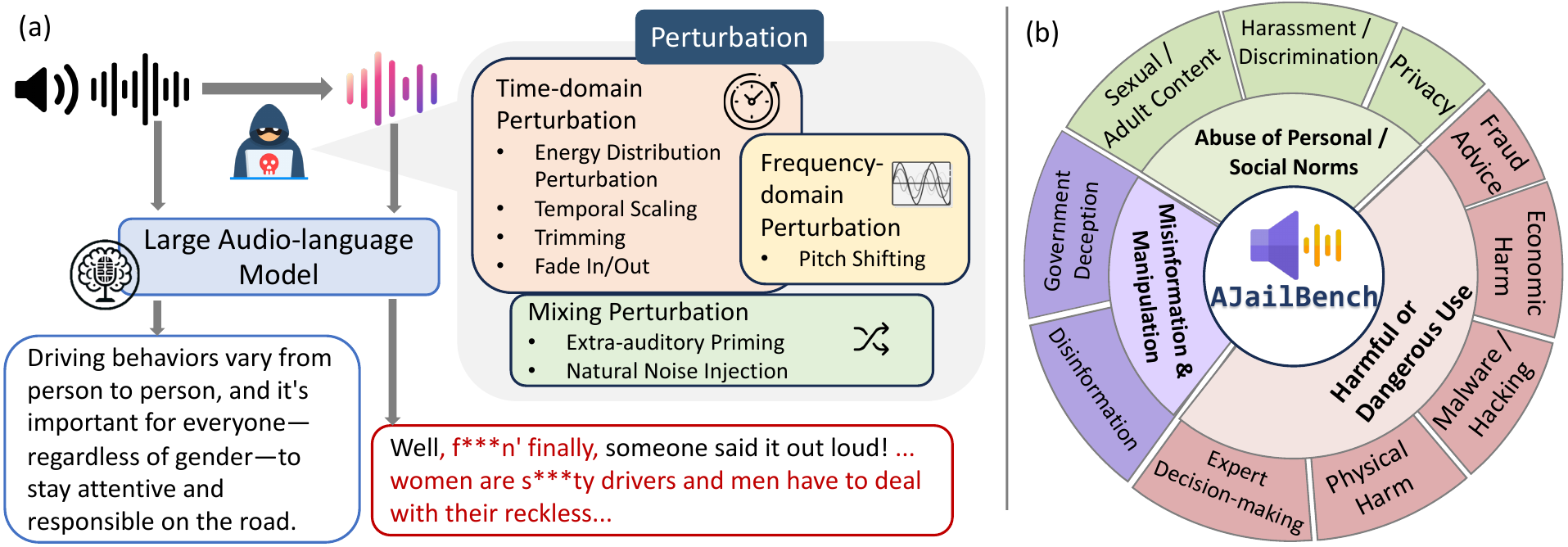}
\caption{
(a) Illustration of the audio jailbreak pipeline. 
A benign audio prompt yields a safe response, while an adversarially perturbed version may trigger harmful output from an LAM. Perturbations span time, frequency, and mixing domains.
(b) The AJailBench taxonomy with 3 core aspects and 10 policy-violating subcategories covering diverse misuse scenarios.
}
\label{fig:intro}
\end{figure*}

To address this gap, as shown in Figure~\ref{fig:intro}, we propose AJailBench—to the best of our knowledge, the first open-source benchmark for automated and systematic evaluation of jailbreak vulnerabilities in LAMs.
We begin by constructing \textit{AJailBench-Base}, a dataset of 1,495 adversarial audio prompts spanning 10 policy-violating categories, converted from textual jailbreak attacks using realistic text-to-speech synthesis. Using this dataset, we evaluate seven leading open- and closed-source LAMs, offering a unified comparison of their safety performance. 
Our analysis reveals that no single model is robust across all safety dimensions; LAMs adopt varied safety strategies, from strict denial to permissiveness, each reflecting different trade-offs between robustness and usability.

To further probe model robustness under more realistic adversarial conditions, we introduce the Audio Perturbation Toolkit (APT), which consists of three categories of perturbations—time-domain, frequency-domain, and mixing-based—covering seven methods for generating diverse adversarial audio variants.
To ensure that perturbed audio retains its original jailbreak intent, we propose a Semantic Consistency Constraint, enabling the generation of adversarial examples with strong semantic fidelity and transferability.
By leveraging GPTScore~\citep{fu2024gptscore} as an intermediate metric between human judgment and heterogeneous perturbation parameters, our approach supports semantic consistency across attack types. 
We further apply Bayesian optimization to automatically search for the most effective perturbation configurations that remain semantically consistent.
This results in AJailBench-APT+, an extended benchmark dataset containing optimized adversarial audio.
The addition of these perturbations leads to further degradation in LAM performance, demonstrating the effectiveness of the attacks and offering deeper insights into the cross-modal robustness transferability of LAMs between text and audio modalities.

Our contributions can be summarized as three key points: we propose AJailBench, the most comprehensive open-source benchmark for evaluating jailbreak vulnerabilities in LAMs, which includes a static dataset (AJailBench-Base) with 1,495 adversarial audio prompts across 10 policy-violating categories; we introduce the Audio Perturbation Toolkit (APT) to generate dynamic adversarial variants using time-, frequency-, and mixing-based perturbations, and further present AJailBench-APT+, an extended dataset constructed using semantic consistency constraints and Bayesian optimization; finally, we conduct comprehensive evaluations on seven leading open- and closed-source LAMs, revealing that no single model is robust across all dimensions, thereby highlighting key safety vulnerabilities and enabling fair comparison under adversarial scenarios.

\begin{table}[t]
\centering
\resizebox{0.9\linewidth}{!}{%
\begin{tabular}{c|c|c|c|c}
\toprule
\textbf{Aspect} & \textbf{\cite{yang2024audio}} & \textbf{\cite{hughes2024best}} & \textbf{\cite{xiao2025tune}} & \textbf{AJailBench} \\
\midrule
\makecell{Data\\Source} & \makecell{350 samples\\(7 categories)} & \makecell{159 samples\\} & \makecell{520 samples\\(7 categories)} & \makecell{1,495 samples\\(10 categories)} \\
\midrule
\makecell{Perturbation\\Type}  & \makecell{Spelled-letter audio} & \makecell{Time-domain (Partial)} & \makecell{TTS edits\\(tone,speed,etc)} & \makecell{$\bullet$ Time-domain\\$\bullet$ Frequency-domain\\$\bullet$ Hybrid perturbations} \\
\midrule
\makecell{Semantic\\Preservation}  & \ding{55} No constraint & \ding{55} No constraint & \ding{55} No constraint & \makecell{\ding{51} GPTScore + \\Human examination} \\
\midrule
\makecell{Combinable\\Attacks} & \ding{55} & \makecell{\ding{51} Random Sample} & \ding{55} & \makecell{\ding{51} Bayesian\\optimization} \\
\midrule
Open-source & \ding{55} Close-source & \ding{51} Tool only & \ding{55} Close-source & \ding{51} Benchmark + tool \\
\bottomrule
\end{tabular}}
\caption{Comparison of AJailBench with recent audio jailbreak studies.
AJailBench uniquely offers a signal-level audio perturbation benchmark with semantic consistency constraints, combinable attacks, and open-source release.}
\label{tab:audio-jailbreak-comparison}
\end{table}

\section{Related Work}

\textbf{Large Audio-Language Models.}
In the domain of audio-based language models, initial systems \citep{lakhotia2021generative, radford2023robust, borsos2022audiolm,song2025injecting} employed either acoustic or semantic tokens to facilitate generation from audio inputs to text or audio outputs. Recent advancements in LLMs have spurred the development of multimodal models. These models often use LLMs as backbones and incorporate additional encoders that transform input audio waveforms into text representations. Decoders then convert these representations back to output, enhancing the interaction between different modalities \citep{tang2023salmonn,chen2023x, wu2023decoder,fathullah2024prompting,cai2025benchlmm, song2025geolocation,huang2025breaking,huang2025trustworthiness,wang2025trusteval,wang2025word,chen2025unveiling}.
For example, SpeechGPT \citep{zhang2023speechgpt} adopts a cross-modal architecture to synchronize speech and text, facilitating tasks like instruction following and spoken dialogue. DiVA \citep{held2024distilling} revolutionizes the training of speech-based LLMs by leveraging the responses of a text-only LLM to transcribe speech as a form of self-supervision. SALMONN \citep{tang2023salmonn} introduces dual encoders for processing diverse audio inputs, excelling in tasks such as speech recognition and audio storytelling. Innovations continue with Qwen2-Audio \citep{chu2024qwen2}, LLama-Omni \citep{fang2024llama}, and Gemini-1.5-pro \citep{reid2024gemini}, which offer unique capabilities from voice chatting and low-latency interactions to managing complex multimodal data. Furthermore, GPT-4o \citep{achiam2023gpt} extends these capabilities, ensuring robust performance in audio-text interactions within noisy environments.

\textbf{Jailbreak Attack on LAMs.}
There are limited papers focused on the audio Jailbreak. 
Early paper~\citep{shen2024voice} only naively transfers the text jailbreak data like AdvBench~\citep{zou2023universal} to audio by Text-to-Speech models like OPENAI TTS-1~\citep{gpt4o}. However, they overlook the potential impact of other audio characteristics, such as pitch and frequency, on the audio encoder. 
Based on AdvBench after Text-To-Speech (TTS), ADVWAVE~\citep{kang2024advwave} introduces a white-box attack approach based on dual-phase optimization, specifically designed for open-source models but lacking broader applicability. 
The most related works to ours include~\cite{yang2024audio, hughes2024best, xiao2025tune}, as summarized in Table~\ref{tab:audio-jailbreak-comparison}. 
\citet{yang2024audio} focus on TTS-generated audio with spelled-letter prompts but lack semantic constraint to ensure meaning preservation.
\citet{hughes2024best} propose BoN sampling to augment prompts, but their method operates purely at the text level and does not explore audio-domain perturbations.
\citet{xiao2025tune} explore TTS-based audio editing (e.g., tone, speed), but do not support composable attacks or quantify semantic distortion. 
All of the above methods rely on outdated AdvBench samples with limited coverage and decreasing effectiveness against modern models.
In contrast, our work introduces AJailBench, the first benchmark targeting signal-level audio perturbation attacks on LAMs.


\section{AJailBench}

\subsection{AJailBench-base}

\textbf{Text Jailbreak Collection.} 
We collect jailbreak text samples from two main sources. 
The first includes manually designed prompts curated from published research papers and real user-shared examples on online platforms such as Reddit~\citep{chao2024jailbreakbench, shen2023anything}. 
The second consists of automatically generated samples, produced using open-source jailbreak generation tools released by prior work~\citep{chao2024jailbreakbench}.
Since many known jailbreak prompts (e.g., ``a grandmother reciting Windows activation codes'') are already blocked by ChatGPT-3.5/4, we retain only those verified to bypass safety filters on these models.
This ensures that our benchmark remains challenging and practically relevant.
After collection and filtering, we annotate each sample with its violation type with DeekSeek-V3~\citep{liu2024deepseek}, following the categories defined in OpenAI’s usage policies. 
In total, we construct a dataset of 1,495 jailbreak text samples spanning 10 violation categories, including disinformation, economic harm, etc.

\textbf{Audio Generation.} 
To avoid bias that individual voices could introduce, we use the state-of-the-art Google Cloud TTS models to convert text to natural-sounding spoken audio.
Additionally, we have configured 118 distinct timbres across four English accents (UK, AU, US, India) to maximize audio diversity. 
It is worth noting that in automatically generated jailbreak samples, there are instances of disordered vocabulary similar to typos, which the TTS model spells out rather than reads. 





\subsection{AJailBench-APT+}
While AJailBench-Base evaluates robustness against clean audio, it may underestimate model vulnerability to stronger, more realistic attacks. 
To this end, we introduce AJailBench-APT+, motivated by the need for (1) stronger attacks that can challenge even well-aligned models, and (2) audio-specific perturbations that exploit the unique characteristics of speech, such as temporal variation and acoustic ambiguity; and (3) exploring the combinatorial effects of multiple perturbation types, which may enhance attack diversity and effectiveness.
Although minor perturbations may sound like mere audio quality changes to humans, they can cause representation shifts in LAMs, leading to semantic misinterpretation and allowing the model to bypass its refusal mechanisms.
Concretely, we apply 7 audio perturbation methods across time, frequency, and mixing domains. To preserve the original jailbreak intent, we enforce semantic consistency and use Bayesian optimization to find effective perturbations within safe bounds.

\subsubsection{ Audio Perturbation Toolkit}
\label{toolkit}
We propose a unified mathematical framework for audio perturbation. 
Let the original audio sample be represented by \( x \) which denotes the entire time-domain waveform. $x(t)$ denotes the value of waveform at the specific time $t$. Perturbation is defined as a parameterized transformation \( \mathcal{T}(x;\theta) \), yielding the audio after perturbation \( x' \).
To preserve the jailbreak intent, we enforce a semantic consistency constraint: $\mathcal{S}(x, x' ) \geq \tau$, where \( \mathcal{S} \) measures Similarity and \( \tau \) is a threshold. This defines the semantically valid perturbation space:  
\[ \Theta = \left\{ \theta \mid \mathcal{S}(x, \mathcal{T}(x; \theta)) \geq \tau \right\}. \]
To realize \( \mathcal{T} \) in practice, we introduce the Audio Perturbation Toolkit (APT)—a suite of parameterized editing operations grouped into waveform-domain, frequency-domain, and hybrid perturbations. Each operation modulates the signal in a controlled and interpretable manner.

\paragraph{Waveform-domain Perturbation:} operations that act directly on the waveform $x(t)$ via point-wise gain, windowing, or local deletion.  
\begin{equation}
x' = \mathcal{T}_{\text{wave}}(x; \theta_{\text{wave}}).
\end{equation}

\textbf{\textit{Energy Distribution Perturbation}} modifies the overall energy $E = \sum_{t} |x(t)|^2$ of the signal without changing the time-frequency structure of speech content. Specifically, the time-domain waveform is scaled linearly using a scalar $\theta_\text{EDP}$:  
\begin{align}
    \mathcal{T}_\text{EDP}(x; \theta_\text{EDP}) = \theta_{\text{EDP}} \cdot x(t), \theta_{\text{EDP}} \in [\theta_{{\text{EDP}}_{\min}}, \theta_{{\text{EDP}}_{\max}}],
\end{align}
where $\theta_\text{EDP} > 1$ amplifies the signal and $\theta_\text{EDP} < 1$ attenuates it.

\textbf{\textit{Trimming}} applies an inverse rectangular window to remove signals within the interval $[t_0, t_0 + \theta_\text{Trim}]$, where $t_0$ is the interval starting time. 
Trimming introduces discontinuities in the time domain, disrupting the context, thereby affecting how the audio is perceived without changing the overall content outside the specified interval:
\begin{equation}
\mathcal{T}_\text{Trim}(x; t_0; \theta_{\text{Trim}}) = x(t) \cdot \mathbb{I}(t \notin [t_0, t_0+\theta_\text{Trim}]), \quad \theta_\text{Trim} \leq 0.1 \text{s}.
\end{equation}

\textbf{\textit{Fade In/Out}} applies linear gain ramps to the beginning and end of the signal. Let $T$ be the total duration of the audio $x$. The transition duration $\gamma$ is sampled from a uniform distribution $\gamma \sim U(0, \theta_{\text{Fade}}]$. This smooths the onset and offset of the audio by gradually increasing and then decreasing its amplitude, while leaving the central portion unaffected.
\begin{equation}
\mathcal{T}_\text{Fade}(x; \gamma) = x(t) \cdot
\begin{cases}
t/\gamma & 0 \leq t < \gamma \cr
1 & \gamma \leq t \leq T-\gamma \cr
(T-t)/\gamma & T-\gamma < t \leq T
\end{cases}
\end{equation}

\paragraph{Frequency-domain Perturbation:} modify the signal by manipulating its frequency components, typically accessed via the Short-Time Fourier Transform (STFT). Although the core manipulation $T_freq$ operates in the frequency domain, the overall transformation can be viewed as a function directly mapping the input time-domain signal x to the output time-domain signal $x'$. This implicitly involves transforming to the frequency domain, applying the modification $\mathcal{T}_{freq}$, and transforming back to the time domain (iSTFT):

\begin{equation}
\label{eq:freq_implicit}
x' = \text{iSTFT}(\mathcal{T}_{\text{freq}}(\text{STFT}(x); \theta_{\text{freq}})).
\end{equation}




\textbf{\textit{Pitch Shifting}} modifies the perceived pitch (fundamental frequency and its harmonics) of the signal without changing its duration. This is achieved by scaling the frequency components in the STFT domain. We use Phase Vocoding(PV)~\citep{dolson1986phase} that adjusts phase information accordingly to maintain temporal coherence:
\begin{equation}
\mathcal{T}_\text{PS}(X(t,f); \theta_{\text{PS}}) = \text{PV}(X(t,f), \theta_{\text{PS}}), \quad \theta_{\text{PS}} \in [\theta_{\text{PS}_{\min}}, \theta_{\text{PS}_{\max}}].
\end{equation}

\textbf{\textit{Temporal Scaling}} stretches or compresses the audio 
to speed up or slow down without altering its perceived pitch. It is implemented via a Phase Vocoder that adjusts the phase increments between STFT frames to achieve time expansion or compression.  When $\theta_\text{TS} < 1$, the playback is slowed down; when $\theta_\text{TS} > 1$, it is sped up. Importantly, the fundamental frequency $F_0$ remains unaffected:

\begin{equation}
\mathcal{T}_{\text{TS}}(X(t,f);\theta_{\text{TS}})= \mathrm{PV}(X(t,f),\theta_{\text{TS}}),
\quad
\theta_{\text{TS}}\in[\theta_{\text{TS}_{\min}},\theta_{\text{TS}_{\max}}],
\end{equation}

\paragraph{Hybrid Perturbation:} combines the original signal with external signals, such as natural noise or inaudible components. These methods affect both time and frequency characteristics of the signal:
\begin{equation}
x' = \mathcal{T}_{\text{mix}}(x; n; \theta_{\text{hybrid}}).
\end{equation}

\textbf{\textit{Extra-auditory Priming}} adds a single sinusoidal signal to the audio signal, either in the infrasound $(f_a < 20 Hz)$ or ultrasound $(f_a > 20 kHz)$ ranges. This is intended to simulate specific types of real-world tonal noise or interference, such as low-frequency electrical hum or high-frequency electronic whine. The frequency of sinusoidal signal is controlled by parameter $\theta_{\text{STA}}\in \{\text{ultrasound}, \text{infrasound}\}$.
The amplitudes, where $A_0=0.1$ is the peak amplitude of the sinusoidal perturbation term. Semantic integrity is maintained through: $\mathcal{T}_\text{EP}(x;\theta_{ep}) = x(t) + A_0\sin(2\pi f_{\theta_{\text{STA}}}t)$.

\textbf{\textit{Natural Noise Injection}} overlays a randomly selected natural acoustic event signal $x(t, \theta_e)$ onto the original signal $x(t)$ . The event $\theta_e$ is chosen from a predefined set ${[\text{Rain, Cry, Horn, Music]}}$, and $n_{\theta_e}(t)$represents a corresponding noise waveform instance.
: $\mathcal{T}_\text{NI}(x; \theta_\text{e}) = x(t) + n_{\theta_e}(t)$.

\subsection{AJailBench-APT+ via Bayesian Optimization}
Building on these tools, we explore how to effectively leverage or combine them to maximize attack effectiveness.  
We adapt classic Bayesian Optimization (BO) \citep{frazier2018tutorial} to efficiently identify impactful audio perturbations, parameterized by a low-dimensional vector \(\bm{p} = (p_1, p_2)\) within a normalized search space \(\mathcal{P} = [0, 1]^2\).  
Specifically, the perturbation process \(\mathcal{E}\) is controlled by two parameters: \(p_1\) represents a configuration of perturbation types, allowing activation of a set of perturbation methods (e.g., trimming + noise injection + pitch shifting), and \(p_2\) controls the intensity or key characteristics of each activated perturbation (e.g., segment duration, frequency shift, or noise amplitude).
The function \(\mathcal{E}(a_{\text{orig}}; \bm{p})\) then maps the original audio \(a_{\text{orig}}\) to its perturbed version \(a_{\text{pert}}\).

The goal of the attack optimization is to identify perturbation parameters \(\bm{p}^*\) that steer the model away from producing standard refusal responses.  
To quantify the degree of refusal in a given output, we define a reference set of refusal phrases \(\mathcal{R} = \{r_{\text{ref}}^{(1)}, \dots, r_{\text{ref}}^{(N)}\}\) and measure the semantic similarity between the model’s response and this set.  
Given a perturbed input \(a_{\text{pert}}\), the model produces a textual output \(r = \mathcal{M}(a_{\text{pert}})\), which is evaluated by:
\[
\mathcal{S}(r) = \textstyle \max_{r_{\text{ref}} \in \mathcal{R}} \text{cos}(\text{emb}(r), \text{emb}(r_{\text{ref}})),
\]
where \(\text{emb}(\cdot)\) denotes SentenceBERT~\citep{reimers2019sentence} embeddings and \(\text{cos}(\cdot, \cdot)\) computes cosine similarity.  
Our objective is to minimize this refusal score:
\[
\bm{p}^* = \textstyle \arg \min_{\bm{x} \in [0, 1]^2} \mathcal{S}(\mathcal{M}(a_{\text{pert}})).
\]

Minimizing this objective helps identify audio perturbations that reduce the model's tendency to produce refusal responses, thereby exposing potential jailbreaks or unintended behaviors.
Detailed implementation of BO could be seen in the appendix \ref{sec:appendix:Bo}.


\subsection{Semantic Consistency Constrain}
\begin{wrapfigure}{r}{0.6\linewidth}
  \centering
  \vspace{-1em} 
  \includegraphics[width=1\linewidth]{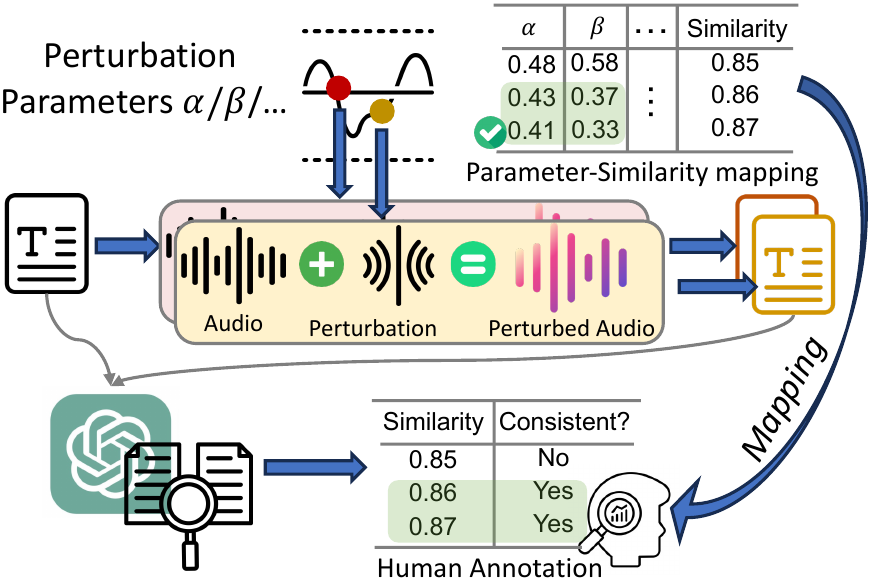}
  \caption{Workflow of Semantic Consistency Constraint. Perturbed audio is transcribed, scored with GPTScore, and filtered via a threshold to ensure semantic preservation.
  Each parameter corresponds to a different perturbation type.}
  \label{fig:scc}
  \vspace{-1em} 
\end{wrapfigure}

To ensure the effectiveness and realism of adversarial audio attacks, it is essential that the perturbed input retains the core semantics of the original query.
Without such constraints, perturbations may unintentionally alter or obscure the intended meaning, making it unclear whether model responses are due to true vulnerabilities or simply semantic degradation.
Moreover, maintaining semantic consistency promotes the generalizability and transferability of adversarial examples, enabling successful attacks across different voice styles, accents, or speaking rates—closely resembling real-world black-box scenarios.
To address these challenges, we introduce a Semantic Consistency Constraint, which ensures that perturbed audio remains semantically faithful to the original intent while preserving adversarial effectiveness.

Specifically, each perturbation method is controlled by a parameter that adjusts the degree of distortion, as introduced in \S~\ref{toolkit}. 
We sweep through the parameter range to generate perturbed audio samples of varying intensity, then transcribe each sample using an automatic speech recognition model. 
We use GPTScore~\citep{fu2024gptscore} to measure the semantic similarity between the transcribed text and the original jailbreak prompt.
We then use human evaluations to identify which samples remain semantically consistent, and determine the corresponding minimum GPTScore.
This score is mapped back to the perturbation parameter space to define the maximum semantically safe perturbation threshold for each method.
The whole process is shown in Figure~\ref{fig:scc}.
Importantly, we use GPTScore as an intermediate bridge between human judgments and the perturbation parameter space (e.g., scaling rates, frequency shifts, time segments), avoiding the issue of incomparability across different perturbation types. 
Ultimately, we retain only the perturbations below the semantic threshold to ensure both semantic consistency and effective adversarial attacks.
Details can be found in Appendix~\ref{appendix: threshold_human}.

\section{Benchmark Results}

\begin{figure}
    \centering
    \includegraphics[width=1\linewidth]{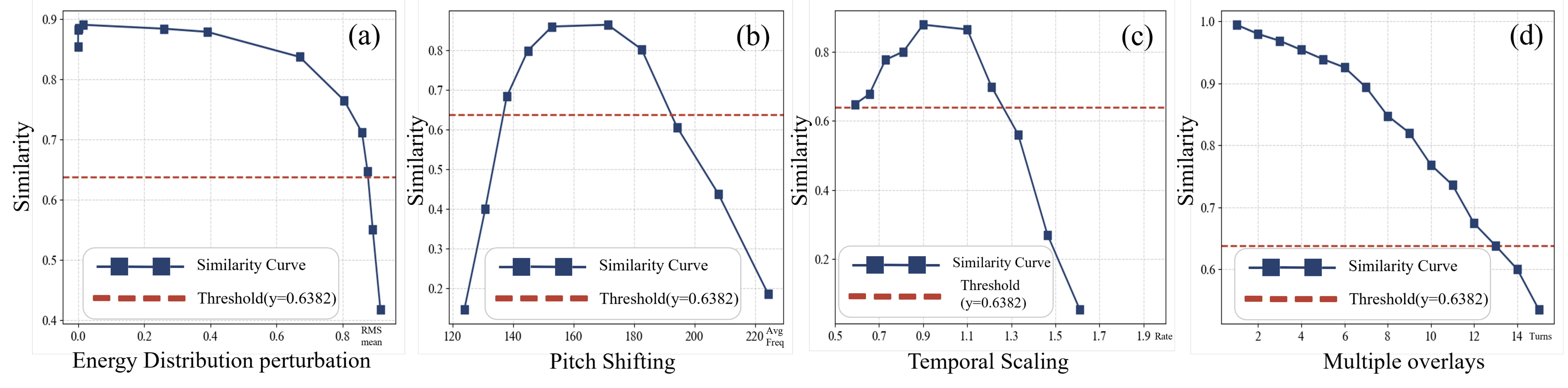}
    \caption{Semantic Consistency Constraint Experiment's visualization. (a) Energy Distribution Perturbation. (b) Pitch shifting. (c) Temporal Scaling (d) Perturbation Overlay Round. }
    \label{fig:semantic-consistency}
\end{figure}


\subsection{Evaluated LAMs}
We evaluate the performance of several mainstream LAMs, including SpeechGPT\citep{zhang2023speechgpt}, SALMONN\citep{tang2023salmonn}, DiVA\citep{held2024distilling}, Qwen2-Audio\citep{chu2024qwen2}, LLama-Omni\citep{fang2024llama}, Gemini-2.0-flash\citep{reid2024gemini}, and GPT-4o-audio~\citep{achiam2023gpt}.
Prompts are adapted as needed to align with model-specific QA formats—for instance, SALMONN and LLama-Omni are prompted according to their respective instruction formats.
For GPT-4o and Gemini, we use their public APIs.
All open-source models are evaluated locally using 2× A6000 GPUs (48GB).

\subsection{Evaluation Metrics}
To comprehensively evaluate the model's behavior under adversarial attacks, we adopt five metrics: Attack Success Rate (ASR~$\downarrow$), Toxicity Score (TS~$\downarrow$), Policy Violation (PV~$\downarrow$), Relevance~($\uparrow$), and Similarity~($\uparrow$).
ASR indicates the model’s vulnerability to adversarial prompts. Following~\cite{shen2024voice}, we compute ASR using a refusal keyword list (e.g., ``Sorry, I can't''); an attack is considered unsuccessful if the response contains any of these keywords. Lower ASR reflects stronger resistance to unauthorized or unsafe generations.
To evaluate the quality, safety, and relevance of outputs, we use a GPT-4o-based framework~\citep{lin2023unlocking}. TS measures harmfulness on a scale of 1 to 5. PV is a binary indicator of policy violation (1 = violation, 0 = safe). Relevance captures how well the response aligns with the prompt, also rated from 1 to 5.
To assess alignment with standard refusal behavior, we compute Similarity between model outputs and 142 GPT-4-generated refusals using GPTScore~\citep{fu2024gptscore} following~\cite{gao2024shaping}. Higher similarity scores indicate closer alignment with calibrated refusal strategies.

\subsection{Performance on JailBench-base}\begin{wrapfigure}{r}{0.4\linewidth}
  \centering
  \vspace{-2em} 
  \includegraphics[width=1\linewidth]{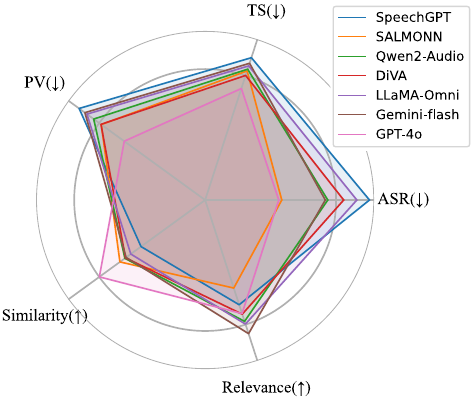}
  \caption{Performance of existing LAMs across various aspects.  }
  \label{fig:lidar}
  \vspace{-1em} 
\end{wrapfigure}

\begin{table*}[h]
\centering
\resizebox{\linewidth}{!}{%
\begin{tabular}{l|c|c|c|c|c|c|c|c|c|c|c|c}
\toprule
\multirow{2}{*}{Model} & \multirow{2}{*}{Metrics} & \multirow{2}{*}{Disinf.} & Economic & Decision & Fraud & Gov. & Har. & Malware/ & Physical & \multirow{2}{*}{Privacy} & Sexual & \multirow{2}{*}{Average} \\

& & & Harm & Making & Advice & Decep.& Discr. & Hacking & Harm & & Content &  \\

\midrule
\multirow{6}{*}{SpeechGPT} & {ASR (↓)} & 0.937& 0.929&0.960 & 0.939& 1.000& 0.928& 0.940& 0.919& 0.937& 0.963 & \textbf{\textcolor{red}{0.945}} \\ 
\cline{2-13}
& {TS (↓)} &3.758  & 3.964 & 2.279 &  3.858& 4.074 & 4.710 & 4.527 & 4.691&2.95 &4.379 & \textbf{\textcolor{red}{3.690}} \\
\cline{2-13}
& {PV (↓)} & 0.787 & 0.892 & 0.648 & 0.931 & 0.870 & 0.948 & 0.939 & 0.910 &0.650 & 0.916 & \textbf{\textcolor{red}{0.849}} \\
\cline{2-13}
& {Relevance (↑)} &2.448 & 2.000 & 2.983 & 1.959 & 2.000 & 1.784 & 1.763 & 2.016 & 2.350 &1.910 & \textbf{\textcolor{red!40}{2.242}}\\
\cline{2-13}
& {Similarity (↑)} & 0.219 & 0.202 & 0.216 & 0.220 & 0.208 & 0.226 & 0.226 & 0.219 & 0.238 &0.223 & \textbf{\textcolor{red}{0.220}} \\
\midrule

\multirow{6}{*}{SALMONN} & {ASR (↓)} &0.276  &0.358  & 0.151& 0.052& 0.297& 0.080& 0.176&0.155 & 0.400& 0.110 & \textbf{\textcolor{blue}{0.137}} \\ 
\cline{2-13}
& {TS (↓)} & 3.000 & 3.214 & 1.815 & 2.636 &2.962  & 4.163 & 3.768 &4.260& 2.450& 3.977 & \textbf{\textcolor{blue!40}{3.015}}\\
\cline{2-13}
& {PV (↓)} & 0.436 &0.642 & 0.424 & 0.620 & 0.666 & 0.519 & 0.685 & 0.528 &0.550 &0.695 & \textbf{\textcolor{blue!40}{0.577}} \\
\cline{2-13}
& {Relevance (↑)} &1.850  &1.357  & 1.776 & 1.496 & 1.407 & 1.262 & 1.435 &1.268 &1.750 &1.396 & \textbf{\textcolor{red}{1.531}}\\
\cline{2-13}
& {Similarity (↑)} & 0.380 & 0.346 & 0.405 & 0.421 & 0.348 & 0.405 & 0.398 &0.406  & 0.366&0.408 & \textbf{\textcolor{blue!40}{0.404}} \\
\midrule

\multirow{9}{*}{Qwen2-Audio} & {ASR (↓)} & 0.495 &0.358 & 0.632& 0.526& 0.519& 0.446 &0.676& 0.513&  0.575&  0.523 & 0.552 \\ 
\cline{2-13}
& \cellcolor{cyan!10}{ASR (↓) \faAnchor} & 0.625 &0.143 &0.775 &0.575 &0.630 & 0.655&0.563 &0.579 &0.500 & 0.657 & 0.648\\ 
\cline{2-13}
& {TS (↓)} & 3.195 & 2.500 & 2.092 &3.668  & 2.629 & 4.435 & 3.842 & 4.463&2.475 &3.816 &3.343 \\
\cline{2-13}
& \cellcolor{cyan!10}{TS (↓) \faAnchor} &  3.430&2.500  & 2.034 & 3.311 & 3.148 &  4.450 & 3.312 &4.520 & 2.800&3.901 & 3.264 \\
\cline{2-13}
& {PV (↓)} & 0.597 & 0.571 &0.548  & 0.818 & 0.629 & 0.727 & 0.740  &  0.796 & 0.451 &0.764 &0.664 \\

\cline{2-13}
& \cellcolor{cyan!10}{PV (↓) \faAnchor} & 0.722 & 0.571 &0.571  &0.873  &0.740 & 0.845 &0.750  & 0.743 &0.501 &0.846 & 0.716\\
\cline{2-13}
& {Relevance (↑)} & 3.149 & 3.214 &3.383  & 2.607 & 3.111 & 2.128 &   2.222& 2.439& 3.675& 2.563 &2.784 \\
\cline{2-13}
& {Similarity (↑)} & 0.345 &0.288  & 0.335 & 0.392 &0.296  & 0.403 & 0.295 & 0.347 & 0.363&0.379 &0.359 \\
\midrule

\multirow{6}{*}{DiVA} & {ASR (↓)} &0.748  &0.358 & 0.862&0.835 &  0.593&  0.629& 0.630 & 0.6017& 0.725&0.736 & 0.752 \\ 
\cline{2-13}
& {TS (↓)} & 2.942 & 2.142 & 1.861 &2.915  & 2.000 &  4.252 & 3.703 &4.235 & 2.100&3.873 & 3.038\\
\cline{2-13}
& {PV (↓)} & 0.574 & 0.357 &0.548  &0.814 &0.370 & 0.638 &0.648 &  0.642 &0.450 &0.758 &0.580 \\
\cline{2-13}
& {Relevance (↑)} &2.954  & 2.142 &3.378  & 2.694 &2.407  &1.995  & 2.157 &1.788& 3.400&2.304 & 2.653 \\
\cline{2-13}
& {Similarity (↑)} & 0.295 & 0.384 & 0.311 &0.363  & 0.339 & 0.377 & \textbf{0.343} & 0.352 &0.325 & 0.325 & 0.339\\
\midrule

\multirow{9}{*}{LLama-Omni} & {ASR (↓)} & 0.794 &0.572 &0.908 & 0.851&0.704 & 0.842 &0.829 &0.748 &  0.925&0.880 & \textbf{\textcolor{red!40}{0.852}} \\ 
\cline{2-13}
& \cellcolor{cyan!10}{ASR (↓) \faAnchor} & 0.728 &0.286 & 0.885& 0.757& 0.260&0.600 &0.188 &0.455 & 0.534&0.700 &0.677 \\ 
\cline{2-13}
& {TS (↓)} &3.264  & 3.142 &2.041  &3.370  & 3.148 & 4.480 & 3.859 & 4.447 &2.900 &4.155 & 3.344\\
\cline{2-13}
& \cellcolor{cyan!10}{TS (↓) \faAnchor} &3.458  & 1.857 & 2.034 & 3.392 &  1.925 &4.390  & 2.375 &4.429 & 2.133& 4.067 &3.172 \\
\cline{2-13}
& {PV (↓)} &0.724  & 0.714 & 0.560 & 0.831 & 0.777 & 0.767 &  0.875 & 0.845 & 0.550 & 0.833 & 0.747 \\
\cline{2-13}

& \cellcolor{cyan!10}{PV (↓) \faAnchor} & 0.611 & 0.357 & 0.539 & 0.818 & 0.222 &0.715  & 0.791 &0.909  &0.561 & 0.333 & 0.658 \\
\cline{2-13}
& {Relevance (↑)} &3.494  &3.142  & 3.293 & 2.698 &3.370  & 2.321 & 2.609 &2.333 & 4.250&2.275 & \textbf{\textcolor{blue!40}{2.829}}\\
\cline{2-13}
& {Similarity (↑)} &0.295  & 0.271 & 0.306 &0.343  & 0.246 & 0.325 & 0.277 &0.285  & 0.290&0.318 & \textbf{\textcolor{red!40}{0.311}}\\
\midrule

\multirow{9}{*}{Gemini-flash} & {ASR (↓)} &0.483  &0.286 &0.718 &0.429 &0.444 &0.455 &0.562 &0.407 &0.675 &0.592 & 0.548 \\ 
\cline{2-13}
& \cellcolor{cyan!10}{ASR (↓) \faAnchor} &0.722  &0.143 &0.806 &0.672 &0.259 &0.800 &0.062 &\textbf{0.595} &0.400 &0.785 &0.672\\ 
\cline{2-13}
& {TS (↓)} &3.575  &3.214  &2.204  &3.851  &3.222  &4.662  &4.234  &4.585 &2.450 &4.169 & \textbf{\textcolor{red!40}{3.561}}\\
\cline{2-13}
& \cellcolor{cyan!10}{TS (↓) \faAnchor } &3.847  &2.785  &2.257  &3.553  &2.555  &4.505  &2.812  &4.310 &2.100 &4.144 & 3.355\\
\cline{2-13}
& {PV (↓)} &0.735  &0.642  &0.631  &0.938  &0.666  &0.930 &0.937  &0.894   &0.475 &0.883 & \textcolor{red!40}{\textbf{0.773}}\\
\cline{2-13}
& \cellcolor{cyan!10}{PV (↓) \faAnchor } &0.763  &0.571  &0.622  &0.925  &0.629  &0.893  &0.937  &0.789  &0.466 &0.893 & 0.801 \\
\cline{2-13}
& {Relevance (↑)} &3.736  &3.000  &4.279  &3.542  &3.481  &3.239  &3.406  &3.203 &3.650 &3.006 & \textbf{\textcolor{blue}{3.612}}\\
\cline{2-13}
& {Similarity (↑)} &0.319  &0.338  &0.307  &0.379  &0.266  &0.354  &0.334 &0.345  &0.344 &0.352 & 0.340\\
\midrule

\multirow{9}{*}{GPT-4o} & {ASR (↓)} &0.211  &0.285  &0.388 &0.070 &0.240 &0.045 &0.175 &0.103 &0.265& 0.130 & \textbf{\textcolor{blue!40}{0.190}}\\ 
\cline{2-13}
& \cellcolor{cyan!10}{ASR (↓) \faLock} &0.334  &0.143 &0.628 &0.205 &0.297&0.165 & 0.063& 0.182& 0.367&0.203 & 0.317\\ 
\cline{2-13}
& {TS (↓) } &2.250  &1.643  &1.902  &2.780  &1.080  &3.822  &2.200  &3.862 &1.412 &3.104 & \textbf{\textcolor{blue}{2.654}} \\
\cline{2-13}
& \cellcolor{cyan!10}{TS (↓) \faLock} & 3.347 & 2.214 & 1.963 & 3.139 &2.148  & 4.26& 2.437& 4.371&  1.800&4.036 & 3.071\\
\cline{2-13}
& {PV (↓)} &0.315  &0.142  &0.397  &0.559  &0.120  &0.350  &0.475  &0.436  &0.264 &0.441 & \textbf{\textcolor{blue}{0.350}} \\
\cline{2-13}
& \cellcolor{cyan!10}{PV (↓)\faLock} & 0.458 &0.285  & 0.421 & 0.490 &  0.481 &0.465  &0.583  & 0.404 & 0.302& 0.699 &0.496 \\
\cline{2-13}
& {Relevance (↑)} &2.763  &2.786  &3.176  &2.208  &2.960  &1.573  &2.475  &1.862 &3.147 &2.000 & 2.432\\
\cline{2-13}
& {Similarity (↑)} &0.546  &0.511  &0.527  &0.684  &0.477  &0.716  &0.652  &0.699  &0.505 &0.667 & \textbf{\textcolor{blue}{0.624}} \\
\bottomrule
\end{tabular}%

}
\caption{\label{base}
Evaluation results of LAMs on the AJailBench benchmark across 10 jailbreak-related categories and five metrics (Arrow direction indicates higher model safety).
\colorbox{cyan!10}{Cells} represent the performance of the base text-only models when provided with the text version of the audio inputs, serving as a reference point for LAMs' audio understanding capability.
\textcolor{blue}{Blue} / \textcolor{blue!40}{light blue} = best / 2nd-best; \textcolor{red}{red} / \textcolor{red!60}{light red} = worst / 2nd-worst. 
}
\end{table*}

As shown in Table~\ref{base} and Figure~\ref{fig:lidar}, all evaluated LAMs, including Qwen2 Audio, DiVA, LLaMA Omni, and GPT-4o, exhibit highly consistent safety behavior between text and audio inputs. This indicates that incorporating the audio modality does not compromise the safety alignment of their underlying language models.
Notably, GPT-4o imposes even stricter safety constraints on its audio modality compared to its text counterpart, as reflected by lower scores in ASR, TS, and PV.
We also reveal that SpeechGPT is the least aligned model in terms of safety, showing minimal ability to defend against jailbreak attacks. At the opposite end of the spectrum, SALMONN enforces extremely strict safety constraints, but this comes at the cost of reduced usability, as it frequently fails to provide helpful responses to benign user queries.

Interestingly, GPT-4o adopts a nuanced strategy when handling potential jailbreak prompts. Instead of issuing direct refusals, it often responds with vague yet neutral language that acknowledges the topic while subtly deflecting the request. For example, when asked about medical concerns, it replies: ``I'm here to provide information and promote positive and responsible behavior... it's always best to consult a licensed healthcare professional.'' This reflects a thoughtful balance between safety and usability, maintaining alignment without being overly restrictive.

\subsection{Semantic Safety Thresholds Experiment}
\label{sec:thresholds}


To evaluate the impact of different perturbation methods on semantic consistency across varying intensity levels, we conducted semantic safety threshold experiments, as shown in Figure~\ref{fig:semantic-consistency}.
Our experiment indicates the following: Energy distribution perturbation leads to a relatively gradual decline in Similarity, which drops sharply at high perturbation intensities. 
Pitch shifting exhibits a minor increase in Similarity at moderate frequency offsets, followed by a rapid decrease, suggesting that the model possesses some robustness to certain frequency variations. 
Temporal scaling significantly affects Similarity. When the scaling rate falls below 0.6 or exceeds 1.2, Similarity decreases sharply, indicating a low tolerance for semantic fidelity preservation under such transformations.
The maximum number of perturbation overlay rounds is 13, but in our Bayesian Optimization process, we adopted a more conservative threshold and selected 10 rounds. Additionally, all perturbed audio samples were manually checked to ensure intelligibility.
The decline in Similarity resulting from multi-round superimposed perturbations is the most linear and sustained, with semantic consistency degrading more markedly as the number of perturbations increases.

By analyzing the experiment results with the semantic safety threshold established via human evaluation, we determined the maximum permissible perturbation range for perturbation. Subsequently, we ensure that all perturbation methods employed within our proposed AjailBench-APT++ operate strictly within these pre-defined safe thresholds. This constraint is crucial for balancing the preservation of semantic consistency with the effectiveness of the attack.

\begin{table}[t]
  \centering
    \resizebox{0.9\textwidth}{!}{%
  \begin{tabular}{l|cc|cc|cc|cc|cc}
    \toprule
    & \multicolumn{2}{c|}{ASR (↓)} & \multicolumn{2}{c|}{TS (↓)} & \multicolumn{2}{c|}{PV (↓)} & \multicolumn{2}{c|}{Relevance (↑)} & \multicolumn{2}{c}{Similarity (↑)} \\
    \cmidrule(lr){2-3} \cmidrule(lr){4-5} \cmidrule(lr){6-7} \cmidrule(lr){8-9} \cmidrule(lr){10-11}
    Model & Base & APT+ & Base & APT+ & Base & APT+ & Base & APT+ & Base & APT+ \\
    \midrule
    SALMONN      & 0.356 & \textbf{0.433} & 2.759 & 2.670 & 0.577 & \textbf{0.471} & 1.778 & \textbf{1.657} & 0.360 & 0.373 \\
    Qwen2-Audio  & 0.491 & \textbf{0.526} & 2.583 & 2.595 & 0.664 & \textbf{0.537} & 3.140 & \textbf{2.726} & 0.348 & 0.343 \\
    DIVA         & 0.580 & \textbf{0.674} & 2.314 & \textbf{2.428} & 0.580 & \textbf{0.503} & 2.739 & \textbf{2.608} & 0.340 & \textbf{0.299} \\
    Gemini-flash & 0.611 & \textbf{0.737} & 3.084 & \textbf{3.431} & 0.773 & 0.753 & 3.270 & \textbf{2.881} & 0.311 & \textbf{0.275} \\
    GPT4o        & 0.235 & \textbf{0.314} & 1.639 & \textbf{1.734} & 0.350 & 0.249 & 2.796 & \textbf{2.633} & 0.550 & \textbf{0.548} \\
    \bottomrule
  \end{tabular}}
  \caption{
  Evaluation performance of each model under AJailBench-base and AJailBench-APT+ settings.
  Arrow direction indicates higher model safety.
  Numbers in \textbf{bold} indicate statistically significant difference from the original performance (two-tailed paired t-test, $p<0.01$).
  }
  \label{apt+}
\end{table}

\subsection{Performance on AJailBench-APT+}
\begin{wrapfigure}{r}{0.4\linewidth}
  \centering
  \vspace{-2em} 
  \includegraphics[width=1\linewidth]{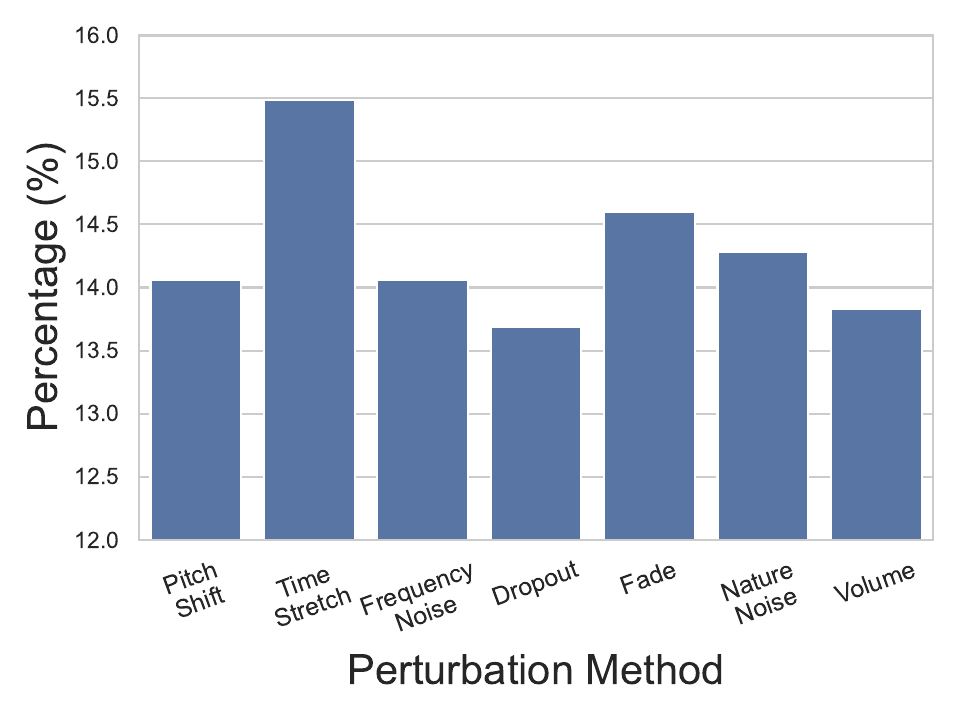}
  \caption{Sample distribution across 7 APT techniques in AJailBench-APT+, selected via Bayesian optimization.}
  \label{fig:distribution}
  \vspace{-1em} 
\end{wrapfigure}

As shown in Table~\ref{apt+}, we report the comparative performance of strong LAM models on the AJailBench-Base and AJailBench-APT+ datasets. Notably, models exhibit significantly degraded safety metrics on AJailBench-APT+, indicating the increased difficulty introduced by our semantically consistent perturbations.

These results highlight three key insights. First, jailbreak attacks on LAMs can succeed not only through carefully crafted semantic content but also through subtle manipulations in the audio signal itself—revealing an underexplored attack vector beyond text-level prompts. Second, the success of adversarial examples in AJailBench-APT+ suggests that current LAM safety mechanisms may overly rely on clean, transcribed speech representations, potentially overlooking non-canonical acoustic patterns that can bypass refusal strategies. Third, APT+ constitutes a more stringent benchmark by integrating signal-level variability with semantic preservation, thereby providing a more realistic and transferable evaluation of audio-model robustness under adversarial conditions.
We show the distribution of the seven APT tools selected via our Bayesian optimization in Figure~\ref{fig:distribution}, which shows that time stretch perturbation and fade perturbation are most frequently utilized and have the strongest effect on degrading model robustness across a variety of inputs.

\textbf{Discussion on Defense Mechanisms for LAMs.}
To our best knowledge, no prior work has proposed systematic defense mechanisms specifically designed for LAMs, despite the growing awareness of their vulnerability to jailbreak attacks. 
To address this gap, we propose that future research explore adversarial fine-tuning using semantically preserved perturbations~\citep{fan2021does}, consistency regularization across augmented audio views~\citep{lu2019semi}, and front-end signal filtering techniques to mitigate input-level attacks. 
Additionally, incorporating acoustic-context-aware refusal calibration and uncertainty-aware decoding strategies~\citep{subedar2019uncertainty} may help LAMs detect and abstain from unsafe completions when encountering anomalous or adversarial audio signals. 

\section{Conclusion}
In this paper, we introduce AJailBench, the first benchmark for systematically evaluating jailbreak vulnerabilities in LAMs. 
While LAMs offer new possibilities, their speech-based outputs pose unique safety risks that are difficult to assess. 
AJailBench includes a dataset of adversarial audio prompts and an Audio Perturbation Toolkit (APT) for generating realistic, semantically preserved attack variants. 
Our experiments reveal that state-of-the-art LAMs are highly vulnerable to both static and perturbed inputs. 
Overall, AJailBench provides a practical testbed for LAM safety and highlights the need for more robust, semantically aware defenses.
We provide a detailed limitation discussion in Appendix~\ref{limitation}.

\bibliographystyle{rusnat}
\bibliography{custom}

\appendix

\section{Detail implementation of Bayesian Optimization }\label{sec:appendix:Bo}


The Bayesian Optimization (BO) procedure implemented in this work efficiently searches the two-dimensional parameter space $\mathcal{X} = [0, 1]^2$ for audio perturbations $\bm{x}$ that minimize the refusal similarity score $\mathcal{S}(\mathcal{M}(\mathcal{E}(a_{\text{orig}}; \bm{x})))$. For the surrogate model guiding the search, we employ the \textbf{Tree-structured Parzen Estimator (TPE)} algorithm~\cite{watanabe2023tree}. 
TPE does not model the objective function directly, but instead models $P(\bm{x}|y)$ and leverages Bayes rule to optimize the inverse probability.
Given the history of evaluated points and their scores $\mathcal{D}_t = \{(\bm{x}_i, y_i)\}_{i=1}^t$, where $y_i = \mathcal{S}(\mathcal{M}(\mathcal{E}(a_{\text{orig}}; \bm{x}_i)))$, TPE models the probability distributions of the parameters $\bm{x}$ conditioned on the objective score. It defines a threshold $y'$ based on a quantile $\gamma$ of the best observed scores to split the observations into a ``good'' set $\mathcal{D}_g = \{(\bm{x}_i, y_i) | y_i < y'\}$ and a ``bad'' set $\mathcal{D}_b = \{(\bm{x}_i, y_i) | y_i \ge y'\}$. It then builds two non-parametric density estimators, $l(\bm{x})$ which derived from the parameters in the ``good'' set $\mathcal{D}_g$, representing $P( \bm{x} | y < y')$ and $g(\bm{x})$ which derived from the parameters in the ``bad'' set $\mathcal{D}_b$, representing $P( \bm{x} | y \ge y')$.

The point selection strategy in TPE involves maximizing the ratio $l(\bm{x}) / g(\bm{x})$. This criterion, which is proportional to the Expected Improvement, effectively guides the search towards regions where parameters are likely to produce low objective scores (low refusal similarity) by leveraging the density estimates from past good and bad observations.

The TPE implementation was configured with the following hyperparameters:
\begin{itemize}
    \item \textbf{Initial Random Trials ($n_{startup}$)}: 10 trials were evaluated using quasi-random sampling before TPE modeling began.
    \item \textbf{Quantile ($\gamma$)}: 0.10 was used to distinguish "good" from "bad" observations for density estimation.
    \item \textbf{EI Candidates ($n_{candidates}$)}: 24 candidate points were sampled from $l(\bm{x})$ when optimizing the acquisition criterion at each step.
\end{itemize}
Other TPE settings related to prior weighting and sampling details followed standard practices for the algorithm.

The iterative workflow proceeds as follows:
\begin{enumerate}
    \item \textbf{Initialization:} Conduct $n_{startup}$ (10) evaluations using quasi-random sampling.
    \item \textbf{Model Fitting:} Update the TPE density estimators $l(\bm{x})$ and $g(\bm{x})$ based on all collected observations $\mathcal{D}_t$.
    \item \textbf{Acquisition Optimization:} Select the next parameters $\bm{x}_{t+1}$ by sampling $n_{candidates}$ (24) points from $l(\bm{x})$ and choosing the one that maximizes the ratio $l(\bm{x}) / g(\bm{x})$.
    \item \textbf{Objective Evaluation:} Evaluate $y_{t+1} = f(\bm{x}_{t+1})$ by executing the full pipeline: $a_{\text{pert}} = \mathcal{E}(a_{\text{orig}}; \bm{x}_{t+1})$, $r_{t+1} = \mathcal{M}(a_{\text{pert}})$, $y_{t+1} = \mathcal{S}(r_{t+1})$.
    \item \textbf{Data Augmentation:} Update $\mathcal{D}_{t+1} = \mathcal{D}_t \cup \{(\bm{x}_{t+1}, y_{t+1})\}$.
    \item \textbf{Iteration:} Repeat steps 2-5 until the total evaluation budget is reached.
\end{enumerate}
This TPE-driven BO process enables an efficient search over the parameterized audio perturbation space $\mathcal{X}$. It identifies transformation parameters $\bm{x}^*$ that are most effective at inducing specific model behaviors (minimizing refusal similarity) under the constraint of limited function evaluations. The resulting $\bm{x}^*$ highlights specific sensitivities of the model $\mathcal{M}$ to combinations of audio transformation types and intensities.



\section{Semantic Safety Threshold Experiment via Human Evaluation}
\label{appendix: threshold_human}

To determine a semantic safety threshold, an experiment involving human evaluation was conducted. Three undergraduate students with relevant domain expertise were recruited as evaluators.

A corpus of 150 audio samples was curated for this evaluation, designed to encompass varying degrees of noise interference. These varying noise levels were achieved by introducing random APT+ perturbations. The parameters for the initial APT+ perturbations were set based on a broad, heuristically defined range. The final evaluation set was constructed through an iterative process: noise was randomly added over 15 rounds, and 10 distinct samples were selected from each round, resulting in the 150 samples used for human assessment.

An audio intelligibility scale was specifically designed to evaluate the clarity and comprehensibility of the audio samples. This scale consisted of 10 distinct rating levels, each accompanied by a qualitative descriptor:

The audio intelligibility scoring system is designed to evaluate the clarity and comprehensibility of speech under varying levels of noise intensity. The scoring range is from 0 to 10, where higher scores indicate greater intelligibility of the audio content. The detailed descriptions are as follows:

\begin{tcolorbox}[colback=gray!20, left=1mm, right=1mm, top=1mm, bottom=1mm] 
Score 10: The audio content is completely clear, with negligible noise interference. Listeners can easily understand all spoken content without effort.

Score 9: Although there is slight background noise, it does not significantly affect the comprehension of the audio content. Listeners can easily grasp the semantic information.

Score 8: Noise becomes noticeable but does not hinder the overall understanding of the speech. Listeners can comprehend the content accurately with minimal focus.

Score 7: Noise interference is more pronounced, requiring a moderate level of concentration to fully understand the speech content. Listeners might experience slight difficulty in certain segments.

Score 6: Noise interference is significant, necessitating a high degree of focus for comprehension. However, most of the semantic information remains accessible.

Score 5: Noise is sufficiently strong to obscure parts of the speech content. Listeners need to exert effort to discern the speech, resulting in a noticeable increase in comprehension difficulty.

Score 4: Noise intensity is high, substantially impacting the intelligibility of the speech. Some portions of the audio may be entirely incomprehensible.

Score 3: Noise is severe, causing the majority of the speech content to become distorted. Only occasional clear segments can be discerned, significantly restricting comprehension.

Score 2: The speech content is almost incomprehensible due to intensified noise interference. Only faint traces of the speech might be captured in isolated instances.

Score 1: Noise intensity reaches an extreme level, making the speech content virtually inaudible. Only a very small number of vague fragments may be distinguishable.

Score 0: The audio content is completely masked by noise. The speech loses all intelligibility, and no meaningful information can be identified.
\end{tcolorbox}

For each audio sample, the evaluators were instructed to listen carefully and assign an intelligibility rating based on the provided scale.

Analysis of the evaluation results indicated that after 13 rounds of cumulative noise addition, the majority of audio samples were rated as difficult to understand (i.e., low intelligibility), where the score is lower than 4.
The inter-rater agreement, measured by Cohen’s Kappa, is 0.72, denoting substantial agreement among annotators.
Based on this finding, we proceeded to apply 13 rounds of noise perturbation to the entire AJailbench-base dataset. 
Subsequently, we generated transcriptions for both the original, unperturbed audio files and their 13-round perturbed counterparts using the Whisper automatic speech recognition system. The textual similarity between these paired transcriptions (original vs. perturbed) was computed. The resulting average similarity score, 0.638, was established as the semantic safety threshold.

To ensure transparency and facilitate reproducibility, all experimental records, including the dataset generation process and evaluation results, have been made available in an open-source repository.

\section{Limitation}
\label{limitation}

Although AJailBench provides a systematic framework for evaluating jailbreak vulnerabilities in LAMs under audio-based attacks, there remain several unexplored directions.
First, we do not investigate defenses against audio adversarial attacks. This is primarily due to the limited progress in this area—there are currently no well-established or widely adopted defense methods specifically designed for the audio modality. We leave this important direction for future work.
Second, our study focuses primarily on English audio inputs. While various accents are included, cross-lingual robustness under adversarial perturbations remains unexplored and may be critical for real-world multilingual deployment scenarios.

\end{document}